\documentclass[pre,aps,reprint,superscriptaddress,showkeys]{revtex4-1}
\usepackage[utf8]{inputenc}
\usepackage{graphicx,booktabs,amsmath,amssymb}
\usepackage[usenames,dvipsnames,table]{xcolor}
\usepackage{url}
\newcounter{comments}

\newcommand{\vek}[1]{\boldsymbol{#1}}
\newcommand{\abs}[1]{\lvert #1\rvert}
\newcommand{\sE}{\mathsf{E}}
\newcommand{\sO}{\mathsf{O}}
\newcommand{\ie}{\emph{i.e.}}
\newcommand{\neff}{n_{\mathrm{eff}}}
\newcommand{\tb}[1]{\textbf{#1}}
\definecolor{lightsteelblue}{RGB}{176,196,222}
\begin{document}
\title{Network-based recommendation algorithms: A~review}
\author{Fei Yu}
\affiliation{Department of Physics, University of Fribourg, 1700 Fribourg, Switzerland}
\author{An Zeng}
\affiliation{Department of Physics, University of Fribourg, 1700 Fribourg, Switzerland}
\affiliation{School of Systems Science, Beijing Normal University, Beijing, P. R. China}
\author{Sébastien Gillard}
\affiliation{Department of Physics, University of Fribourg, 1700 Fribourg, Switzerland}
\author{Matúš Medo}
\email{matus.medo@unifr.ch}
\affiliation{Department of Physics, University of Fribourg, 1700 Fribourg, Switzerland}

\noaffiliation

\begin{abstract}
Recommender systems are a vital tool that helps us to overcome the information overload problem. They are being used by most e-commerce web sites and attract the interest of a broad scientific community. A recommender system uses data on users' past preferences to choose new items that might be appreciated by a given individual user. While many approaches to recommendation exist, the approach based on a network representation of the input data has gained considerable attention in the past. We review here a broad range of network-based recommendation algorithms and for the first time compare their performance on three distinct real datasets. We present recommendation topics that go beyond the mere question of which algorithm to use---such as the possible influence of recommendation on the evolution of systems that use it---and finally discuss open research directions and challenges.
\end{abstract}

\maketitle

\section{Introduction}
The rapid development of the Internet has a great impact on our daily lives and has significantly changed the ways in which we obtain information. Movie fans, instead of going to a physical shop to buy or rent a DVD, can now use one of the many online movie-on-demand or rental services to watch the movie they want. Online services have similarly simplified our access to books and music. The same thing happens to our social lives: instead of going to bars to meet with old and possibly also new friends, we now have multiple online social networks which allow us to communicate with friends as well as to find new ones. However, the convenience brought by the Internet comes with the burden to choose from the immense number of possibilities---which movie to watch, which song to hear, whose Tweets to read---which has become to known as the information overload problem~\cite{maes1994agents}.

Since it is often impossible for a person to evaluate all the available possibilities, the need has emerged for automated systems that would help to identify the potentially interesting and valuable candidates for any individual user. Many information filtering techniques have been proposed to meet this challenge~\cite{hanani2001information}. One representative method is the search engine which returns the most relevant web pages based on the search keywords provided by the users~\cite{langville2011google}. Though effective and commonly used, search engines have two main drawbacks. First, they require the users to specify the keywords describing the contents that they are interested in, which is often a difficult task, especially when one has little experience with a given topic or, even, when one does not know what they are looking for. Second, search results are not personalized which means that every user providing the same keywords obtains the same results (this problem can be corrected by assessing the individual's history of searches). This is crucial because the tastes and interests of people are extraordinary diverse and ignoring them is likely to lead to inferior filtering performance.

The second class of information filtering techniques, recommender systems, overcomes the above-mentioned problems. The goal of a recommender system is to use data on users' past interests, purchase behavior, and evaluations of the consumed content, to predict further potentially interesting items for any individual user~\cite{resnick1997recommender}. These data typically take form of ratings given by users to items in an integer rating scale (most often 1 to 5 stars where more stars means better evaluation) but it can also be of so-called unary kind where a user is connected with an item only if the user has purchased, viewed, or otherwise collected. Since user tastes and interests are included in the input data, recommendations can be obtained without providing any search queries or keywords. The choice of items for a given user builds on the user's past behavior which ensures that the recommendation is personalized. However, the degree of personalization can be harmed by excessive focus on recommendation accuracy~\cite{mcnee2006being,zhang2008avoiding}.

Collaborative filtering is perhaps the most usual approach to recommendation~\cite{goldberg1992using,schafer2007collaborative}. User-based collaborative filtering evaluates the similarity of users and recommends items that have been appreciated by users who are similar to a target user for whom the recommendations are being computed (analogously, item-based approach builds on evaluating the similarity of items). Other techniques include content-based analysis~\cite{pazzani2007content},
spectral analysis~\cite{goldberg2001eigentaste}, and latent semantic models and matrix factorization~\cite{hofmann2004latent,koren2009matrix}. The last-mentioned class of algorithms has recently gained popularity because of contributing importantly to the winning solution~\cite{koren2010collaborative} in the Netflix prize contest~\cite{bennett2007netflix}. See \cite{su2009survey,koren2011advances,lu2012recommender,tang2013social} for a current review of various aspects of the field of recommendation.

While most recommender systems act on data with ratings, unary data without ratings are the basis for a class of physics-inspired recommendation algorithms. These algorithms represent the input data with a bipartite user-item network where users are connected with the items that they have collected (more information on complex networks and their use for analyzing and modeling real systems can be found in~\cite{albert2002statistical,boccaletti2006complex,newman2010networks}). Classical physics processes such as random walk~\cite{zhou2007bipartite} and heat diffusion~\cite{zhou2010solving} can be then employed on the network to obtain recommendations for individual users. See~\cite{medo2013network} for a review of the basic ideas in network-based recommendation and ranking algorithms. Many variants and improvements of the originally proposed algorithms have been subsequently published and their scope has been extended to, for example, the link prediction problem~\cite{zhou2009predicting,lu2011link} and the prediction of future trends~\cite{zeng2013trend}.

In this review, we select a comprehensive group of recommendation algorithms that act on unary data and compare them for the first time using several recommendation performance metrics and various datasets that differ in their basic properties such as size and sparsity. After introducing the algorithms and the evaluation procedure in Section~\ref{sec:methods}, we present the results in Section~\ref{sec:results}. In this section, we focus in particular on evaluating the contribution of additional parameters that are used by most of the algorithms to improve their performance and make it possible to adjust the algorithm to a particular dataset. In Section~\ref{sec:beyond}, we discuss several questions that are not directly related to recommendation algorithms. In particular, we expand considerably the findings presented in~\cite{gualdi2013crowd,vidmer2015prediction} that can be used to further improve accuracy and diversity of recommendations by limiting the number of users to whom each individual item can be recommended. Finally in Section~\ref{sec:discussion}, we summarize the main conclusions of this review and outline the major research directions for the future.

\section{Methods}
\label{sec:methods}
In this section, we describe the notation, benchmark datasets, recommendation methods, and the evaluation procedure and metrics that are used in this review.

\subsection{Data and notation}
\label{sec:data}
The input data for a recommender system typically consists of past activity records of users. We confine ourselves to the simplest case where the past record for each user is represented by the set of items collected by this user. Further information, such as the time when individual items have been collected or personal information about the user (gender, age, nationality, and so forth) is not required. The input data can be effectively represented by a bipartite user-item network where a user and item node are connected when the corresponding user has collected the given item. In the case when users also rate the collected items, we represent with links only those collected items whose rating is greater or equal than a chosen threshold rating.

The number of user and item nodes in the network is $U$ and $I$, respectively. The total number of links in the network is $L$. In mathematical notation, we speak of the bipartite graph $G(\mathsf{U}, \mathsf{I}, \mathsf{L})$ where $\mathsf{U},\mathsf{I},\mathsf{L}$ is the set of users, items, and links, respectively, and $U:=\abs{\mathsf{U}}$, $I:=\abs{\mathsf{I}}$, $L:=\abs{\mathsf{L}}$. To improve the clarity of our notation, we use Latin letters $i,j$ to label user nodes and Greek letters $\alpha,\beta$ to label item nodes. The degree of user $i$ and item $\alpha$ are labeled as $k_i$ and $k_{\alpha}$, respectively. The degree of a user represents the number of items collected by the user and the degree of an item represents the number of users who have collected the item. Since each link is attached to one user node and one item node, it holds that $\sum_i k_i = \sum_{\alpha} k_{\alpha} = L$.

We evaluate recommendation methods on three distinct datasets.  The Movielens dataset has been obtained from \url{www.grouplens.org}. It consists of 100,000 ratings from 943 users on 1,682 movies (items) with the integer rating scale from 1 (worst) to 5 (best). To obtain unary data, we neglect all ratings below 3 and represent all remaining ratings with user-item links. After this thresholding, the dataset contains 82,520 links. The Netflix dataset has been obtained from \url{www.netflixprize.com}. From the original data, we have chosen 10,000 users and 6,000 movies at random. After the same thresholding procedure as before, there are 701,947 links present. The Amazon data has been originally obtained by crawling the website \url{www.amazon.com} in summer 2005 (see~\cite{slanina2005referee} for details). We further reduced the subset of the Amazon data used in~\cite{zeng2013information} by choosing at random 10,000 users who have given rating 3 or above at least once and are thus not left without links by the thresholding procedure and keeping all the links connected with them. The resulting dataset contains 57,037 links between 10,000 users and 24,403 items. Table~\ref{tab:datasets} summarizes basic characteristics of the three datasets. The most noteworthy difference between them is that while the Movielens and Netflix data are rather dense (at least for the field's standards), the Amazon data are 100-times less dense, which makes it challenging for recommendation algorithms.

\begin{table*}
\centering
\begin{ruledtabular}
\begin{tabular}{lrrrrrrrr}
Dataset   &    $U$ &    $I$ &     $L$ &           density & $E(k_i)$ & $M(k_i)$ & $E(k_{\alpha})$ & $M(k_{\alpha})$\\
\hline
Movielens &    943 &  1,682 &  82,520 & $5.2\cdot10^{-2}$ &  $88$ & 1,018 &  $49$ & 1,116\\
Netflix   & 10,000 &  6,000 & 824,802 & $1.4\cdot10^{-2}$ &  $82$ & 2,120 & $137$ & 9,018\\
Amazon    & 10,000 & 24,403 &  57,037 & $2.3\cdot10^{-4}$ & $5.7$ &   679 & $2.3$ & 1,198\\
\end{tabular}
\end{ruledtabular}
\caption{Basic properties of the used datasets: the number of users $U$, the number of items $I$, the number of links $L$, data density $L / (UO)$, average user degree $E(k_i)$, maximal user degree $M(k_i)$, average item degree $E(k_{\alpha})$, and the maximal item degree $M(k_{\alpha})$.}
\label{tab:datasets}
\end{table*}

\subsection{Recommendation methods}
In this review, we focus on recommendation methods that directly build on a network representation of the input data. The first method of this kind, a probabilistic spreading algorithm~\cite{zhou2007bipartite}, is based on a simple random walk process on the user-item network. This process is first used to compute link weights in a bipartite network projection $G(\mathsf{U}, \mathsf{I}, \mathsf{L})\to G(\mathsf{W})$ where $\mathsf{W}$ is the matrix of link weights in the projected network. In particular the item-projection (\ie, a projection on a monopartite network where only the item nodes remain) is then exploited to compute item recommendation scores for an individual user. This elementary approach has been since then many times modified and generalized in order to improve the accuracy and diversity of the resulting recommendations. We now introduce the original method and a wide variety of its variants that are compared in this review.

\begin{figure*}
\centering
\includegraphics[scale=1]{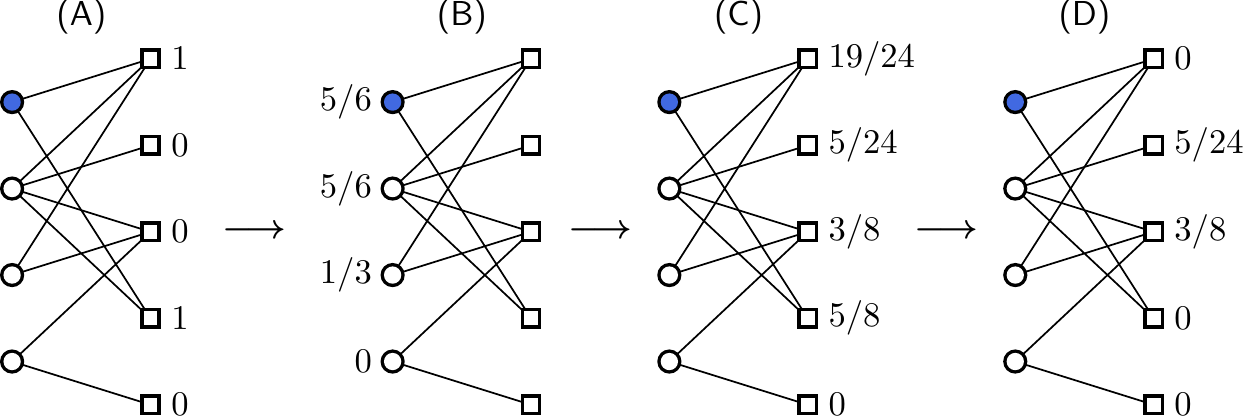}
\caption{An illustration of the Probabilistic spreading algorithm (ProbS). Circles and squares represent users and items, respectively. The color-marked circle represents the target user for whom the recommendation is being computed. (Adapted from~\cite{zhou2010solving}.)}
\label{fig:ProbS}
\end{figure*}

\paragraph*{Probabilistic spreading (ProbS).}
This recommendation method builds on a random walk process on the user-item network~\cite{zhou2007bipartite}. For a given user $i$, we first initially set the unit amount of resource on all items collected by this user; all other items have zero initial resource value (panel A in Figure~\ref{fig:ProbS}). In the first random walk step, the resource spreads from the item side to the user side. Since links of the bipartite network are unweighted, it is natural to assume that the amount of resource on an item node $\alpha$ is divided uniformly into $1/k_{\alpha}$ parts and transmitted to the user side over the network (panel B). In the second random walk step, the resource value on a user node is again divided uniformly and spreads over the network from the user side back to the item side (panel C). The resource amounts on the item side present the final item scores which decide whether an item is recommended to the given user $i$ or not (the higher the score, the better). Note that the items already connected with user $i$ generally have non-zero score (in fact, their scores tend to be the highest of all). To avoid recommending these items to the user, one first sets the score of all items connected with user $i$ to zero (panel D) and then sorts the items according to the resulting score in descending order. A small number of top-ranking items then comprise the final recommendation for the given user.

To obtain a mathematical formulation of the ProbS method, we denote the initial resource vector for user $i$ as $\vek{f}^{(i)}$ and its elements as $f_{\alpha}^{(i)} = a_{i\alpha}$. Since the resource redistribution process is linear, the final resource values $h_{\alpha}^{(i)}$ can be written as
\begin{equation}
\label{ProbSgeneral}
h_{\alpha}^{(i)} = \sum_{\beta=1}^I W_{\alpha\beta}f_{\alpha}^{(i)} = (\mathsf{W}\vek{f}^{(i)})_{\alpha}
\end{equation}
Elements of the redistribution matrix $\mathsf{W}$ follow directly from the description of the probabilistic spreading process as
\begin{equation}
\label{ProbSW}
W_{\alpha\beta} = \frac1{k_{\beta}}\sum_{j=1}^U \frac{a_{j\alpha}a_{j\beta}}{k_j}.
\end{equation}
Here the sum over terms proportional $a_{i\alpha}a_{i\beta}$ corresponds to paths going from $\beta$ to $\alpha$ in two steps through any one of the users, and the division with $k_{\beta}$ and $k_i$ corresponds to the uniform division of resources between all the links originating at nodes $\beta$ and $i$, respectively. The matrix $\mathsf{W}$ represents the item-side projection of the bipartite network. Note that this matrix is asymmetric. If there are some users who have collected both a popular item $\alpha$ and a little popular item $\beta$, then $W_{\alpha\beta}>W_{\beta\alpha}$ because $k_{\alpha} > k_{\beta}$. This is the formal expression of the fact that the users who have collected $\beta$ frequently collects also $\alpha$, but the converse is not true---out of the many users who have collected item $\alpha$, only a few (at most $k_{\beta}$) have also collected item $\beta$. The matrix $\mathsf{W}$ is column normalized as which stems from the fact that it represents a conserving random walk which preserves the total amount of resource.

We now remark briefly on the implementation of ProbS. While the method is mathematically conveniently represented by the matrix $\mathsf{W}$ that aggregates the input bipartite network and can act on the initial network of resources representing the collection of any individual user, it is actually preferable to compute the recommendation scores by implementing the two-step spreading process. The latter approach is obviously advantageous memory-wise: Instead of holding the $I\times I$ matrix $\mathsf{W}$ in memory, one constructs the relevant contribution directly from the input data that comprises $\eta UI$ entries (here $\eta$ is the data density, \ie, the fraction of links between user and item nodes that are actually present). Since data density is usually very small, in online systems it ranges from $10^{-2}$ to $10^{-4}$ or even less, the straightforward matrix representation of the bipartite network with a $U\times I$ matrix of zeros and ones is wasteful and one should use a sparse matrix representation where only an array of links is stored in memory. To effectively realize the spreading process, it is advantageous to prepare both user- and item-sorted array of links which facilitates to quickly determine to which items and users, respectively, the resource spreads. The memory usage can be further halved if in the user-sorted array, for example, only the item identifiers are present instead of the usual user-item pairs. To make it possible to identify the user corresponding to an array entry, it is then sufficient to remember the element at which each user's collection starts. If we know, for example, that a user has collected 20 items and this user's starting element is 10, then the items collected by this user are stored in elements from 10 to 29. Due to data sparsity, computing the recommendation scores using the two-step spreading process is also typically faster than using the matrix $\mathsf{W}$. The same implementation considerations hold for all the methods that we review here.

One of the oldest approaches to recommendation, user-based collaborative filtering, computes recommendation for a user by aggregating over the opinions of all other users whereby giving higher weight whose past ratings are similar to the past ratings of the given user. In the context of data without ratings represented by a bipartite user-item network, user similarity can be computed by the so-called cosine similarity $s_{ij}=\sum_{\alpha=1}^I a_{i\alpha}a_{j\alpha} / \sqrt{k_i k_j}$ where the numerator counts the number of items collected by both user $i$ and user $j$ and the denominator provides a plausible normalization factor (see~\cite{lu2011link} for an extensive review of other similarity metrics in bipartite graphs). The computed score of item $\alpha$ from the point of view of user $i$ is then
\begin{equation}
x_{i\alpha} = \frac{\sum_{j=1}^U s_{ij} a_{j\alpha}}{\sum_{j=1}^U s_{ij}} = (\mathsf{T}\vek{h}^{(i)})_{\alpha}
\end{equation}
where again $h_{\alpha}^{(i)} = a_{i\alpha}$ and the elements of the matrix $\mathsf{T}$ are $T_{ij} = s_{ij} / \sum_{j=1}^U s_{ij}$. The score $x_{i\alpha}$ is structure-wise identical with that presented by Eq.~(\ref{ProbSgeneral}) which tells us that ProbS is intrinsically a form of user-based collaborative filtering. Yet, ProbS features an advantage that has been exploited by its numerous generalizations: It is based on a specific multi-step process which gives us multiple handles to influence the overall behavior of the resulting method and thus accommodate diverse characteristics that one might require from a recommendation method under various circumstances.

\paragraph*{Heterogeneous initial configuration method (Zhou, 2008).}
In the original ProbS vector, the initial resource vector for user $i$ is $f^{(i)}_{\alpha} = a_{i\alpha}$. By summing it over all users, we obtain an estimation of the total power given to item $\alpha$ in the recommendation and obviously $\sum_{i=1}^U f^{(i)}_{\alpha} = k_{\alpha}$; the item's power is proportional to its popularity. One of the first generalizations of the ProbS method has questioned this and asked whether lowering the power of popular items is not beneficial for the resulting recommendation performance~\cite{zhou2008effect}. In particular, the initial resource vector has been suggested in the form
\begin{equation}
f^{(i)}_{\alpha} = a_{i\alpha}k_{\alpha}^{\theta}
\end{equation}
where $\theta$ is a free parameter and $\theta=0$ gives the ProbS method. The total power of item $\alpha$ is then $k_{\alpha}^{1+\theta}$; when $\theta<0$, popular items have less power than in the original ProbS method. The authors have shown that this modification improves the ranking score $r$ which is minimized at $\theta\approx -0.8$ for the studied data set. As one might expect, it's particularly the ranking score of little popular items that is improved markedly when $\theta$ is negative. By contrast, the ranking score of popular items is little sensitive to $\theta$ when $\theta\gtrsim -0.8$ and then quickly deteriorates which explains why the overall optimal value for $\theta$ is close to $-0.8$. Note that at $\theta=-0.8$, the total power of items is proportional to $k_{\alpha}^{0.2}$, \ie, it increases very slowly with item popularity. It has been further reported that choosing a negative value of $\theta$ lowers the average degree of the recommended items and increases the average Hamming distance between the users' recommendation lists.

\paragraph*{Elimination of redundant correlations (Zhou, 2009).}
As noted before, repetitions of the spreading process obtained by applying higher powers of $\mathsf{W}$ on the resource vector do not bring satisfactory results. In~\cite{zhou2009accurate}, the authors propose to combine the scores from the usual probabilistic spreading with the two-step probabilistic spreading as follows
\begin{equation}
h^{(i)}_{\alpha} = \big((\mathsf{W} + \eta\mathsf{W}^2)\vek{f}^{(i)}\big)_{\alpha}
\end{equation}
where $\eta$ is a hybridization parameter. Instead of computing $\mathsf{W}^2$, this is best achieved by actually applying the ProbS spreading scheme twice. Parameter values that optimize the resulting ranking score are typically negative what the authors interpret as a sign of redundant correlations being eliminated by subtracting $\eta\mathsf{W}^2$ from $\mathsf{W}$ (see~\cite{zhou2009accurate} for a more detailed discussion).

\paragraph*{Unequal resource allocation method (Run-Ran, 2010).}
One of the early generalizations of the ProbS method is based on the assumption that the attraction of a node in the redistribution process is proportional to $k^{\theta}$ where $k$ is the node degree and $\theta$ is a tunable parameter~\cite{liu2010personal}. The fraction of resource that is transmitted from item $\beta$ to user $i$ is thus proportional to $a_{i\beta}k_i^{\theta}$ (step 1) and the fraction of resource that is transmitted from user $i$ to item $\alpha$ is proportional to $a_{i\alpha}k_{\alpha}^{\theta}$. The transmission matrix thus takes the form
\begin{equation}
W_{\alpha\beta} = \frac{k_{\alpha}^{\theta}}{\sum_{l=1}^U a_{l\beta}k_l^{\theta}}
\sum_{j=1}^U \frac{a_{j\alpha}a_{j\beta}k_j^{\theta}}{\sum_{j=1}^U a_{j\beta} k_j^{\theta}}
\end{equation}
which simplifies to Eq.~(\ref{ProbSW}) when $\theta=0$ (\ie, when the resource is divided evenly between the neighboring nodes). When $\theta>0$, popular nodes become more attractive and score better; when $\theta<0$, the converse is true. According to the authors, substantial recommendation improvements are observed when $\theta$ is negative~\cite{liu2010personal}. Note that the power terms $k_i^{\theta}$ are rather slow to compute because $\theta$ is in general a floating-point number. To speed up the computation, one can prepare a table with the values $\{1^{\theta}, 2^{\theta},\dots, k_{\mathrm{max}}^{\theta}\}$ where $k_{\mathrm{max}}$ is the highest occurring degree value and then look in the table instead of computing the corresponding term. The same holds for many of the following methods.

\begin{figure*}
\centering
\includegraphics[scale=1]{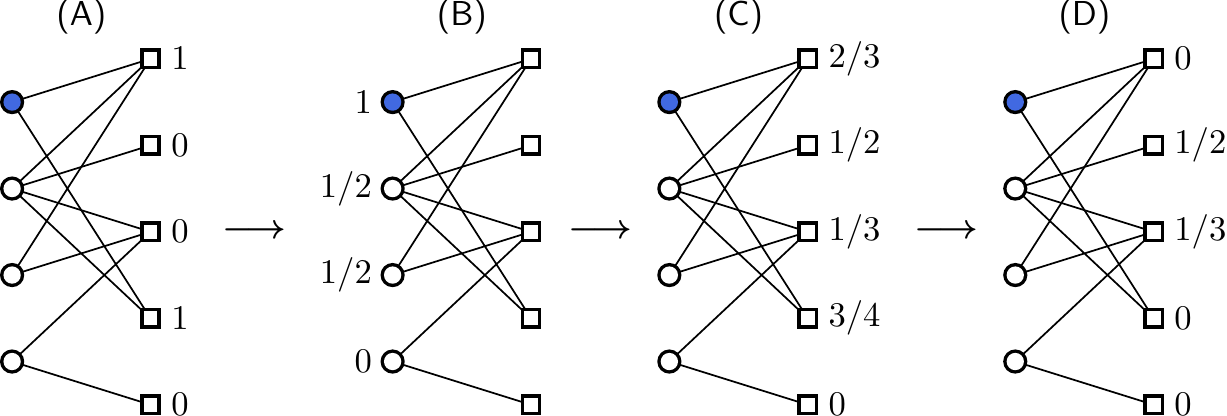}
\caption{An illustration of the Heat spreading algorithm (HeatS). Circles and squares represent users and items, respectively. The color-marked circle represents the target user for whom the recommendation is being computed. (Adapted from~\cite{zhou2010solving}.)}
\label{fig:HeatS}
\end{figure*}

\paragraph*{Heat spreading method (HeatS) and a ProbS-HeatS hybrid method (Zhou, 2010).}
Structure of the Heat spreading algorithm is very similar to that of ProbS with only one fundamental difference. While ProbS relies on dividing the resource uniformly and transmitting it to the other side of the bipartite graph, HeatS is based on an averaging process where a node's score is obtained by averaging over the score of all the nodes it is connected with (see an illustration in Figure~\ref{fig:HeatS})~\cite{zhou2010solving}. Mathematically speaking, the spreading process is now represented by the matrix
\begin{equation}
W_{\alpha\beta}' = \frac1{k_{\alpha}}\sum_{j=1}^U \frac{a_{j\alpha}a_{j\beta}}{k_j}.
\end{equation}
Item scores are computed as before, \ie, $h_{\alpha}^{(i)} = \mathsf{W}'\vek{f}^{(i)}$. Elements of the initial resource vector can be now interpreted as temperature values for individual item nodes and consequently elements of $\vek{h}$ are the resulting temperature values; the ``hot'' item nodes are then recommended to a given user. Note that a similar view has been used before to devise a recommendation method~\cite{zhang2007heat} which however does not build on a bipartite representation of the data and we thus do not discuss it here further. Further discussion on the relation between the probabilistic and heat spreading and, for example, heat spreading and the probability of absorption in sinks in random walk can be found in~\cite{stojmirovic2007information,medo2013network}.

Albeit the difference between $\mathsf{W}$ and $\mathsf{W}'$ is seemingly minor ($\mathsf{W}'$ is row-normalized as opposed to the column-normalized $\mathsf{W}$), it has far-reaching consequences. While the ProbS method favors popular items (the probabilistic spreading process is cumulative and thus an item improves its chances to score high by having many links), HeatS favors little popular items (the heat spreading process is averaging and thus an item improves its chances to score high by having a few links to users with a large resource value). This popularity bias is visible also in Figures~\ref{fig:ProbS} and \ref{fig:HeatS} where the highest ProbS score $3/8$ is achieved by the comparatively well connected third item from the top but the highest HeatS score is achieved by the second item from the top which only has degree one.

We further demonstrate the HeatS's preference for low-degree items and the resulting exceptional diversity of the recommended items in the Results section. However, the items chosen by the method are typically so little popular and peculiar, that the method itself is rendered useless---the corresponding recommendation accuracy is low and, furthermore, it is known that the users actually appreciate some level of familiarity in their recommendation lists as it helps them to build trust in the system's inner working~\cite{swearingen2002interaction}. The observation that ProbS produces satisfactory accuracy but favors popular items and thus lacks in diversity and HeatS produces high diversity but fails in accuracy has motivated a hybrid method which uses an additional parameter $\lambda$ to smoothly interpolate between ProbS and HeatS as the two respective extremes~\cite{zhou2010solving}. From the various functional forms that fulfill this criterion, the best results are produced with
\begin{equation}
\label{PHhybrid}
W_{\alpha\beta}^H = \frac1{k_{\alpha}^{1-\lambda}k_{\beta}^{\lambda}}\sum_{j=1}^U \frac{a_{j\alpha}a_{j\beta}}{k_j}.
\end{equation}
where $\lambda=0$ recovers the HeatS method, and $\lambda=1$ recovers ProbS. Results presented in~\cite{zhou2010solving} show that this combination of two methods goes substantially beyond a mere interpolation between them. By tuning the hybridization parameter $\lambda$ appropriately, the accuracy of ProbS can be maintained or even improved whilst considerably improving recommendation diversity. The optimal value of the hybridization parameter depends on the choice of the objective function (do you simply want to maximize recommendation accuracy or are you perhaps willing to sacrifice some accuracy to further enhance recommendation diversity?) and the data set used.

\paragraph*{Self-avoiding forward and backward propagation (Blattner, 2010).}
Together with~\cite{zhou2010solving}, \cite{blattner2009b} is the first work where the complementary process of probabilistic spreading---heat diffusion (in the paper referred to as ``backward propagation'')---is used in recommendation. One first constructs the transition matrix
\begin{equation}
\label{BRank_matrix}
P_{\alpha\beta}=\frac{1-\delta_{\alpha\beta}}{k_{\beta}}\sum_{j=1}^U \frac{a_{j\alpha}a_{j\beta}}{k_j - 1}.
\end{equation}
With respect to ProbS, there is an important conceptual difference here: $\mathsf{P}$ describes ``non-lazy'' diffusion in which $P_{\alpha\alpha}=0$ (after one diffusion step, the resource cannot end in the initial node). This is achieved by the term $1-\delta_{\alpha\beta}$ with the Kronecker delta function. Since for user $i$, there are now only $k_i-1$ possible items reachable in one step, the normalization is now with $k_i-1$ instead of $k_i$ used for ProbS.

The B-Rank algorithm then computes the score of items in so-called forward and backward propagation for user $i$ as $\vek{h}^{(i)}_F = \mathsf{P}\vek{f}^{(i)}$ and $\vek{h}^{(i)}_B = \mathsf{P}^{\mathrm{T}}\vek{f}^{(i)}$, respectively. Unlike for the ProbS-HeatS hybrid method, these two independent sets of scores are combined non-parametrically by element-wise multiplication; the resulting score of item $\alpha$ is thus $(h^{(i)}_F)_{\alpha}(h^{(i)}_B)_{\alpha}$. Note that for the sake of consistence with the rest of this review, the transition matrix defined by Eq.~(\ref{BRank_matrix}) is already transposed unlike in~\cite{blattner2009b} where it is not and the forward and backward propagation are achieved by applying $\mathsf{P}^{\mathrm{T}}$ and $\mathsf{P}$, respectively. We have further simplified the original notation by omitting user weights (which are set to one anyway) and avoiding the introduction of link weights $h_{i\alpha}$ by directly substituting their explicit form $h_{i\alpha}=a_{i\alpha} / \sqrt{k_i-1}$.

\paragraph*{Biased heat spreading method (Liu, 2011).}
As noted before, the main problem of the heat spreading method HeatS is that it prefers little popular items too strongly. As opposed to the ProbS-HeatS hybrid, \cite{liu2011information} solves this problem by modifying the HeatS spreading matrix alone in the following way
\begin{equation}
W_{\alpha\beta} = \frac1{k_{\alpha}^{\theta}}\sum_{j=1}^U \frac{a_{j\alpha}a_{j\beta}}{k_j}
\end{equation}
where the original HeatS method is recovered with $\theta=1$ and $\theta<1$ enhances the standing of popular items. Up to a missing factor $1/k_{\beta}^{1-\theta}$, this is identical with Eq.~(\ref{PHhybrid}) and this manifests itself in results achieved with this method being similar to those achieved with the ProbS-HeatS hybrid.

\paragraph*{Weighted heat conduction method (Liu, 2011*).}
To compensate the HeatS's preference for little popular items, one may also assign higher weight to links connecting high-degree nodes. The simplest link weight form, $e_{i\alpha}=(k_ik_{\alpha})^{\theta}a_{i\alpha}$, is used in~\cite{liu2011informationB} to modify the heat spreading matrix into the form
\begin{equation}
W_{\alpha\beta} = \frac1{k_{\alpha}}\sum_{j=1}^U \frac{e_{j\alpha}e_{j\beta}}{k_j}.
\end{equation}
Note that this is not entirely consistent because link weights are used as a direct replacement for elements of the adjacency matrix as opposed to, as one might prefer for principal reasons, distributing the amount of resource proportionally to $e_{i\alpha}$ (in which case normalization terms with a sum of $e_{i\alpha}$ over all links from a given node would appear). When $\theta=1$, the distribution of edge weights is too broad because the two input distributions---that of user and item degree, respectively---are broad as well and this decreases the effective number of network links that are used in the recommendation process. The value of $\theta$ that yields the best ranking score is therefore expected to be close to zero which is confirmed by~\cite{liu2011informationB} where they find the optimal $\theta\approx0.14$ for the same Movielens data set as we use here, for example.

\paragraph*{Preferential diffusion method (L\"u, 2011) and its hybrid with (Zhou, 2008).}
To compensate the ProbS's bias towards popular items, \cite{lu2011information} proposes to promote the standing of little popular items in the second diffusion step (from users to items) by making their score proportional to $k_{\alpha}^{\varepsilon}$. The diffusion matrix then reads
\begin{equation}
\label{Wpreferential}
W_{\alpha\beta} = \frac1{k_{\beta}k_{\alpha}^{-\varepsilon}}\sum_{j=1}^U \frac{a_{j\alpha}a_{j\beta}}{\sum_{\gamma=1}^I a_{j\gamma}k_{\gamma}^{\varepsilon}}
\end{equation}
where the denominator can be written as $k_j E_{\gamma\in \mathsf{C}_j}(k_{\gamma}^{\varepsilon})$ where $\mathsf{C}_j$ is the set of items collected by user $j$. When $\varepsilon=0$, this simplifies to the diffusion matrix of ProbS. When $\varepsilon<0$, the score of little popular items increases in comparison with the $\varepsilon=0$ case. The authors note that the intention of this method is the same as that of heterogeneous initial resource method~\cite{zhou2008effect}. Motivated by the considerable difference between the methods, they propose a hybrid method where the initial resource of object $\beta$ is proportional to $k_{\beta}^{\theta}$ and resource redistribution is governed by the $\mathsf{W}$ above. In the tables with results, we label this hybrid methods as ``L\"u, 2011*''.

\paragraph*{Hybrid method with a degree-dependent hybridization (Qiu, 2011).}
The ProbS-HeatS hybrid with a unique hybridization parameter $\lambda$ allows to tune between preferring low-degree items (when $\lambda$ is close to zero and HeatS prevails) and high-degree items (when $\lambda$ is close to one and ProbS prevails). The situation can be made even more leveled by adjusting the chances of low- and high-degree items simultaneously. To do so, \cite{qiu2011item} proposes a hybridization parameter which depends on the degree of the item where spreading starts in the form
\begin{equation}
\lambda = (k_{\beta} / k_{\mathrm{max}})^{\theta}
\end{equation}
where $\theta$ is a free parameter and $k_{\mathrm{max}}$ is the largest item degree in the input data. When $\theta>0$, low-degree items are used to recommend other low-degree items (because the corresponding $\lambda$ is low) and simultaneously high-degree items are used to recommend other high-degree items. When $\theta=0$, $\lambda=1$ for all starting items and the ProbS method is recovered. The authors find that the overall ranking score $r$ can be improved by setting $\theta>0$ and, in particular, the low-degree items benefit as performance metrics computed specifically for them increase significantly. Since the number of items of degree $10$ and less in users' recommendation lists is higher than for other studied methods, the authors claim that their method contributes to solving the cold-start problem.

\paragraph*{A user-based version of HeatS (Guo, 2012).}
As opposed to the majority of network-based recommendation algorithms which effectively build on item-item similarity, network representations of data can be also used to compute user-user similarity and consequently compute recommendation scores in a user-based fashion. Assuming that $s_{ij}$ is the similarity of users $i$ and $j$, the recommendation score of a yet-uncollected item $\alpha$ for user $i$ (hence $a_{i\alpha}=0$) takes the form
\begin{equation}
f^{(i)}_{\alpha} = \sum_{j=1}^U s_{ji}a_{j\alpha}
\end{equation}
The additional normalization with $\sum_{j\neq i} s_{ji}$ that is used in~\cite{liu2009improved}, for example, is actually unnecessary because it only rescales the scores for a given user and thus does not alter the ranking of individual items. In~\cite{guo2012heat}, the authors compare a host of user similarity metrics including two metrics that are direct analogs of the spreading matrices behind ProbS and HeatS. Finally, they also propose a generalized form of the HeatS analog which reads
\begin{equation}
s_{ij} = \frac1{k_i}\sum_{\alpha=1}^I\frac{a_{i\alpha}a_{j\alpha}}{k_{\alpha}^{\theta}}
\end{equation}
where the exponent varies between $0$ and $4$ in the original article.

\paragraph*{ProbS-HeatS method with heterogeneous initial resource (Liu, 2012).}
As the Heterogeneous initial configuration method has shown, ProbS profits from assigning the items with the initial resource values depending on the item degree~\cite{zhou2008effect}. The goal of \cite{liu2012heterogeneity} is the same with the core method being the ProbS-HeatS hybrid as opposed to the pure ProbS before. The initial resource is also assumed in the same form, \ie, $f_{\alpha}^{(i)} = a_{i\alpha}k_{\alpha}^{\theta}$. This method thus has two parameters: The ProbS-HeatS hybridization parameter $\lambda$ and the initial resource heterogeneity parameter $\theta$.

\paragraph*{Modified heat diffusion method (Qiu, 2013).}
Another method that aims to improve the precision and recall specifically for low-degree items has been presented in~\cite{qiu2013heterogeneity}. Its redistribution matrix taking the form
\begin{equation}
W_{\alpha\beta} = \frac1{k_{\alpha}^{\theta}k_{\beta}}\sum_{j=1}^U \frac{a_{j\alpha}a_{j\beta}}{k_j}
\end{equation}
which is similar with both the originally proposed ProbS-HeatS hybrid as well as the Biased heat spreading method~\cite{liu2011information} that the authors cite as a direct inspiration for their work.

\paragraph*{Semi-local diffusion for sparse datasets (Zeng, 2013).}
The extreme sparsity featured by some of the datasets---in some cases, only $10^{-4}$ or less of all possible links are actually present---poses a challenge to all recommendation methods. In particular spreading-based recommendation methods suffer from the fact that when the data is sparse, the two-step resource spreading assigns non-zero score to only a small number of items, especially when the given user is little active. A semi-local diffusion where the spreading process from items through users back to items is repeated a number of times has been considered before~\cite{zhou2010solving} but it has been found unnecessary as in dense networks it generally offers little improvement at high computational cost. Mathematically speaking, such repeated iterations can be represented with
$$
\vek{h}^{(i)} = \mathsf{W}^n\vek{f}^{(i)}
$$
where $\vek{f}^{(i)}$ is the initial resource vector for user $i$ and $n$ is a small integer which represents how many times the basic spreading process is repeated; Eq.~(\ref{ProbSgeneral}) is recovered when $n=1$. Semi-local diffusion with up to repetitions has been studied in \cite{zhou2009accurate}.

As shown in~\cite{zeng2013information}, semi-local diffusion is vital for sparse datasets where it reduces the set of items with zero score and thus directly improves the ranking score. Since the effects of semi-local spreading depend on the degree of the respective user and item, it is advisable to use it in a heterogeneous way. The proposed item-based version has the form
\begin{equation}
h_{\alpha}^{(i)} = (\mathsf{W}\vek{f}^{(i)})_{\alpha} + \sum_{m=2}^n k_{\alpha}^{-\theta} (\mathsf{W}^m \vek{f}^{(i)})_{\alpha}
\end{equation}
where $\mathsf{W}$ is the usual ProbS redistribution matrix and the higher-order contributions are assigned on the basis of the target item degree (it actually turns out that the optimal value of $\theta$ is negative which means that semi-local diffusion is particularly important for evaluating the popular items). The authors of \cite{zeng2013information} suggest that $n=3$ (\ie, considering at most three spreading iterations) provides a compromise between improving the ranking score and recall.

\paragraph*{Similarity-preferential diffusion (Zeng, 2014).}
As we said before, the spreading process effectively assumes an asymmetric item similarity metric. An analogous approach can be used to argue that the ProbS method actually relies on the user similarity metric
\begin{equation}
s_{ij} = \sum_{\beta=1}^I \frac{a_{i\beta}a_{j\beta}}{k_{\beta}}
\end{equation}
which is symmetric and non-negative. The resulting ProbS recommendation score can be then cast in the form $h_{\alpha}^{(i)} = \sum_{j=1}^U a_{j\alpha}s_{ij}/k_j$. This form has the advantage of making it possible to either enhance or suppress the weight of the most similar users as done in~\cite{zeng2014information} where the recommendation score is computed as
\begin{equation}
h_{\alpha}^{(i)} = \sum_{j=1}^U \frac{a_{j\alpha}s_{ij}^{\theta}}{k_j}
\end{equation}
which simplifies to ProbS when $\theta=1$. It is then natural to apply the same thinking to the hybrid ProbS-HeatS method. This gives us
\begin{equation}
s_{ij} = \sum_{\beta=1}^I \frac{a_{i\beta}a_{j\beta}}{k_{\beta}^{\lambda} k_j^{1-\lambda}},\quad
h_{\alpha}^{(i)} = \sum_{j=1}^U \frac{a_{j\alpha}s_{ij}^{\theta}}{k_j^{\lambda}k_{\alpha}^{1-\lambda}}
\end{equation}
which is the actual method that we include in this survey.

\paragraph*{Three-parameter heterogeneous methods (3hybrid1 and 3hybrid2).}
To conclude, we propose two new hybrid methods that combine the previously described methods via three hybridization parameters. The first method, 3hybrid1, is based on the elimination of redundant correlations~\cite{zhou2009accurate} and the preferential diffusion~\cite{lu2011information} with the additional manipulation of resources on the user side. In terms of the spreading process, the initial resource on the item side is split evenly among the neighboring users where the intermediate user resource $s_i$ is raised to the power of $\theta$. The resource is then divided among the neighboring items proportionally to $k_{\alpha}^{\varepsilon}$, thus giving rise to the first set of item scores $f_{\alpha}^{(1)}$. Those values are divided evenly and transmitted to the user side where the power of $\theta$ is again applied on the user resource. Preferential spreading proportional to $k_{\alpha}^{\varepsilon}$ then yields the second set of item scores $f_{\alpha}^{(2)}$. The final item score is obtained as $f_{\alpha}^{(1)} + \eta f_{\alpha}^{(2)}$.

The second method, 3hybrid2, combines heterogeneous initial configuration~\cite{zhou2008effect}, heterogeneous preferential diffusion~\cite{lu2011information}, and the elimination of redundant correlations~\cite{zhou2009accurate}. The initial resource is set to $a_{i\alpha}k_{\alpha}^{\theta}$ and then propagated according to the preferential diffusion method with parameter $\varepsilon$. We thus obtain the first set of item scores $f_{\alpha}^{(2)}$ which is then transformed in the second set of item scores $f_{\alpha}^{(2)}$ by applying preferential diffusion again (note that in the second step, item degree does not further modify the item resource which is propagated by preferential diffusion). The final item score is again obtained as $f_{\alpha}^{(1)} + \eta f_{\alpha}^{(2)}$.

\subsection{Recommendation evaluation}
After having described the considered recommendation methods, we now proceed by detailing the recommendation evaluation process and metrics that are the standardly used for this task~\cite{herlocker2004evaluating,shani2011evaluating}.

\subsubsection{Double and triple division}
\label{sec:triple}
The simplest approach to evaluate a recommendation method is based on a double division of the data into a so-called training set $\sE^T$ and a probe set $\sE^P$. There are various ways how to split the data into $\sE^T$ and $\sE^P$. We use here the simplest and often used random split under which the training set contains 90\% of the original data and the probe set contains the remaining 10\% (other common proportions are 80-20 and 50-50). Since isolated items cannot be recommended to users through the algorithms considered in this paper, we make sure that there are no items of zero degree in the training set. The training set is used as the actual input data for the evaluated method, the obtained recommendation results are compared with the probe data, and the match between them is quantified using some of the evaluation metrics presented in the following section. To eliminate the statistical variation that stems from the randomness of the training-probe division, the results are averaged over multiple splits.

The described approach becomes problematic for recommendation methods with parameters. As we stated in the Introduction, the optimal parameter values usually substantially depend on the input dataset and the metric which is being optimized. Furthermore, there is often no parameter setting or rule that can be used at any time without markedly harming the recommendation performance. If the method parameters are optimized directly by comparing with the probe set, we may get optimistically biased view of the method's performance (\emph{in-sample} estimate). This problem can be overcome with triple division where a small part of the original data (again, 10\% is an often used proportion that we also use here) is moved into a so-called learning set $\sE^L$. We again make sure that no items in the training set are left with zero degree after this division. After computing recommendations on $\sE^T$, method parameters are optimized by assessing the recommendation performance on $\sE^L$. The learned parameter values are then fixed, recommendations are again computed on $\sE^T\cup\sE^L$, and finally the recommendation performance is evaluated by comparing recommendations with $\sE^P$ (\emph{out-of-sample} estimate). Note that while introducing an additional parameter to a recommendation method is bound to improve or keep intact the method's performance in the usual training-probe data division, this is not necessarily the case for the triple data division evaluation. Besides avoiding the optimism bias and over-fitting, triple division thus makes it possible to compare methods that differ in their number of parameters. Results presented in Section~\ref{sec:results} are all based on triple division evaluation which is an important contribution as most of the method-proposing papers do not use this practice.

\subsubsection{Evaluation metrics}
Accurate recommendation means that the recommendation method can effectively identify the items that the target user likes. The primary accuracy metric that we use here is the \emph{ranking score}. It measures whether the ordering of items by the recommendation method matches the users' real preference. For a target user $i$, all uncollected items are ranked by the recommendation score in a descending order; we thus obtain a so-called recommendation list for user $i$. For every user-item pair $(i,\alpha)$ in the probe set, we compute the ranking of item $\alpha$ in $i$'s recommendation score lits, $r_{i\alpha}$, and normalize this quantity with the score list length $I-k_i$ which is simply given by the number of items that have not yet been collected by user $i$. The ranking score $r$ is obtained by averaging the normalized ranking over all probe entries as
\begin{equation}
r = \frac{1}{\abs{\sE^P}}\sum_{(i,\alpha)\in \sE^P} r_{i\alpha}.
\end{equation}
Small $r$ means that the probe items---which, recall, have been actually collected by the respective users---are ranked highly by the given recommendation method which thus can be said to produce accurate recommendations.

When recommendation is used in practice, only a limited number of recommended items are shown to a user by default. This is an important motivation for using evaluation metrics that unlike the ranking score focus only on the top ranked items. The most usual metric of this kind is \emph{precision}. When in the top $L$ places of a user's recommendation list, there are $d_i(L)$ probe items corresponding to user $i$, precision is computed as
\begin{equation}
P_i(L)=\frac{d_i(L)}{L}.
\end{equation}
The overall precision $P(L)$ is obtained by averaging the precision values over all non-isolated users (\ie, users with at least one entry in the probe). Precision thus represents the average fraction of ``hits'' in top $L$ places of all recommendation lists. The higher the fraction, the more accurate the recommendation. We use $L=50$ for this and all the following metrics that depend on the recommendation list length $L$.

It has been documented that excess focus on accuracy can be detrimental to the resulting recommendations, in particular their benefit for the users~\cite{mcnee2006being}. Diversity of recommendations has gradually emerged as an important additional aspect of recommendation evaluation~\cite{herlocker2004evaluating}. More specifically, recommendation diversity refers to how well the recommendation algorithm can uncover users' very personalized preferences, in particular for the fresh (little popular) items. To this end, we employ two kinds of diversity measurement: \emph{personalization} and \emph{novelty} in this paper.

The personalization metric considers how different are the recommendation lists of different users. The difference is usually measured by the Hamming distance. Denoting the number of common items in the top-$L$ place of the recommendation lists of user $i$ and $j$ as $C_{ij}(L)$, the corresponding Hamming distance can be calculated as
\begin{equation}
D_{ij}(L) = 1 - \frac{C_{ij}(L)}{L}.
\end{equation}
This is a number between $0$ and $1$, which respectively correspond to the cases where $i$ and $j$ have the same or entirely different recommendation lists. By averaging $D_{ij}(L)$ over all pairs of users, we obtain the mean Hamming distance $D(L)$ which is referred to as personalization. Apparently, the more the recommendation lists differs from each other, the higher the $D(L)$.

The intrasimililarity measures the overall level of similarity of items in each user's recommendation list. Denoting the set of top $L$ items in the recommendation list of user $i$ as $\sO_L^i$, the intrasimilarity for user $i$ can be written as
\begin{equation}
I_i(L) = \frac{1}{L(L-1)}\sum_{\substack{\alpha,\beta\in\sO_L^i\\
\alpha\neq\beta}} s_{\alpha\beta}
\end{equation}
where $s_{\alpha\beta}$ is a suitable item similarity metric. The usual choice here is the cosine similarity of the items' collection patterns~\cite{zhou2009accurate}, that is
\begin{equation}
s_{\alpha\beta} = \frac1{\sqrt{k_{\alpha}k_{\beta}}}\sum_{i=1}^U a_{i\alpha}a_{i\beta}.
\end{equation}
By averaging $I_i(L)$ over all users, we obtain the mean intrasimilarity of the users' recommendation lists. Here low intrasimilarity is advantageous because it implies that the individual sets of items recommended to users are internally heterogeneous.

Finally, novelty measures the average degree of the items in the top $L$ positions of all recommendation lists. This can be written as
\begin{equation}
N(L) = \frac1{LU} \sum_{i=1}^U\sum_{\alpha\in\sO_L^i} k_{\alpha}
\end{equation}
A small value of $N(L)$ (as compared by a value achieved by a different recommendation method, for example) indicates that novel and comparatively little popular items are being chosen by the given recommendation method.

To summarize, a well-performing recommendation method results in low ranking score, high precision, high personalization, low intrasimilarity, and low novelty score. We shall see in the next section that while there is no method which scores best in all these aspects, there are methods that achieve a better trade-off between accuracy and diversity than the others.

\begin{table*}
\centering
\begin{ruledtabular}
\begin{tabular}{lrrrrrrr}
Method               & Params &   $r$ &$P(L)$ &$R(L)$ &$I(L)$ &$D(L)$ & $N(L)$\\
\hline
\rowcolor{lightsteelblue}
HeatS                &      0 & 0.134 & 0.023 & 0.137 & \tb{0.055} & \tb{0.856} & \tb{25}\\
ProbS                &      0 & 0.094 & 0.073 & 0.476 & 0.354 & 0.617 & 231\\
Blattner, 2010       &      0 & 0.118 & 0.080 & 0.529 & 0.330 & 0.725 & 204\\
ProbS-HeatS          &      1 & 0.074 & 0.085 & 0.539 & 0.303 & 0.808 & 172\\
Liu, 2011            &      1 & 0.075 & 0.083 & 0.530 & 0.300 & 0.805 & 172\\
Guo, 2012            &      1 & 0.099 & 0.071 & 0.460 & 0.354 & 0.618 & 230\\
Qiu, 2013            &      1 & 0.075 & 0.085 & 0.541 & 0.318 & 0.828 & 168\\
Liu, 2011b           &      1 & 0.079 & 0.084 & 0.526 & 0.321 & 0.845 & 158\\
Zhou, 2009           &      1 & 0.071 & 0.088 & 0.562 & 0.324 & 0.790 & 186\\
Zhou, 2008           &      1 & 0.089 & 0.076 & 0.489 & 0.342 & 0.669 & 221\\
L\"u, 2011           &      1 & 0.072 & 0.087 & 0.552 & 0.299 & 0.824 & 167\\
Run-Ran, 2010        &      1 & 0.097 & 0.078 & 0.485 & 0.301 & 0.804 & 179\\
Qiu, 2011            &      1 & 0.088 & 0.074 & 0.485 & 0.350 & 0.630 & 228\\
Zeng, 2013           &      1 & 0.094 & 0.073 & 0.476 & 0.354 & 0.617 & 231\\
Liu, 2012            &      2 & 0.071 & 0.087 & 0.549 & 0.298 & 0.832 & 165\\
L\"u, 2011*          &      2 & 0.071 & 0.087 & 0.551 & 0.295 & 0.832 & 165\\
Zeng, 2014           &      2 & 0.071 & 0.088 & 0.558 & 0.312 & 0.839 & 165\\
\rowcolor{lightsteelblue}
3hybrid1             &      3 & \tb{0.068} & \tb{0.090} & \tb{0.572} & 0.310 & 0.817 & 174\\
3hybrid2             &      3 & 0.070 & 0.089 & 0.566 & 0.314 & 0.808 & 178\\
\end{tabular}
\end{ruledtabular}
\caption{Evaluation of the recommendation methods on the Movielens dataset ($L=50$). Here and in the two following tables, we use bold font to mark the best performance in each metric and shaded rows to mark the methods that are best in at least metric.}
\label{tab:ml}
\end{table*}

\begin{table*}
\centering
\begin{ruledtabular}
\begin{tabular}{lrrrrrrr}
\toprule
Method               & Params &   $r$ &$P(L)$ &$R(L)$ &$I(L)$ &$D(L)$ & $N(L)$\\
\midrule
\rowcolor{lightsteelblue}
HeatS                &      0 & 0.114 & 0.001 & 0.021 & \tb{0.004} & \tb{0.798} & \tb{14}\\
ProbS                &      0 & 0.044 & 0.055 & 0.423 & 0.366 & 0.425 & 2,365\\
Blattner, 2010       &      0 & 0.045 & 0.058 & 0.449 & 0.357 & 0.485 & 2,288\\
ProbS-HeatS          &      1 & 0.040 & 0.061 & 0.470 & 0.337 & 0.567 & 2,140\\
Liu, 2011            &      1 & 0.043 & 0.060 & 0.464 & 0.327 & 0.578 & 2,098\\
Guo, 2012            &      1 & 0.051 & 0.052 & 0.408 & 0.366 & 0.399 & 2,382\\
Qiu, 2013            &      1 & 0.042 & 0.063 & 0.480 & 0.327 & 0.637 & 2,008\\
Liu, 2011b           &      1 & 0.044 & 0.064 & 0.482 & 0.315 & 0.691 & 1,877\\
Zhou, 2009           &      1 & 0.037 & 0.064 & 0.483 & 0.344 & 0.589 & 2,139\\
Zhou, 2008           &      1 & 0.042 & 0.056 & 0.428 & 0.356 & 0.483 & 2,295\\
L\"u, 2011           &      1 & 0.037 & 0.062 & 0.473 & 0.306 & 0.617 & 1,972\\
Run-Ran, 2010        &      1 & 0.044 & 0.055 & 0.422 & 0.366 & 0.424 & 2,366\\
Qiu, 2011            &      1 & 0.044 & 0.054 & 0.416 & 0.366 & 0.412 & 2,377\\
Zeng, 2013           &      1 & 0.044 & 0.055 & 0.423 & 0.366 & 0.425 & 2,365\\
Liu, 2012            &      2 & 0.036 & 0.064 & 0.480 & 0.294 & 0.678 & 1,884\\
L\"u, 2011*          &      2 & 0.037 & 0.062 & 0.472 & 0.301 & 0.631 & 1,951\\
\rowcolor{lightsteelblue}
Zeng, 2014           &      2 & \tb{0.036} & \tb{0.068} & \tb{0.499} & 0.300 & 0.718 & 1,826\\
3hybrid1             &      3 & 0.037 & 0.062 & 0.476 & 0.327 & 0.578 & 2,094\\
3hybrid2             &      3 & 0.037 & 0.064 & 0.480 & 0.343 & 0.582 & 2,149\\
\bottomrule
\end{tabular}
\end{ruledtabular}
\caption{Evaluation of the recommendation methods on the Netflix dataset ($L=50$).}
\label{tab:nf}
\end{table*}

\begin{table*}
\centering
\begin{ruledtabular}
\begin{tabular}{lrrrrrrr}
Method               & Params &   $r$ &$P(L)$ &$R(L)$ &$I(L)$ &$D(L)$ & $N(L)$\\
\hline
\rowcolor{lightsteelblue}
HeatS                &      0 & 0.366 & 0.002 & 0.056 & 0.075 & \tb{0.994} & \tb{3}\\
ProbS                &      0 & 0.362 & 0.004 & 0.135 & 0.055 & 0.990 & 11\\
Blattner, 2010       &      0 & 0.363 & 0.003 & 0.100 & 0.096 & 0.976 & 5\\
ProbS-HeatS          &      1 & 0.361 & 0.004 & 0.141 & 0.037 & 0.964 & 19\\
Liu, 2011            &      1 & 0.361 & 0.004 & 0.138 & 0.028 & 0.942 & 22\\
Guo, 2012            &      1 & 0.362 & 0.005 & 0.150 & 0.037 & 0.986 & 12\\
Qiu, 2013            &      1 & 0.362 & 0.004 & 0.136 & 0.058 & 0.956 & 19\\
Liu, 2011b           &      1 & 0.365 & 0.003 & 0.090 & 0.099 & 0.977 & 4\\
Zhou, 2009           &      1 & 0.253 & 0.004 & 0.145 & 0.053 & 0.984 & 14\\
Zhou, 2008           &      1 & 0.362 & 0.004 & 0.137 & 0.049 & 0.989 & 11\\
L\"u, 2011           &      1 & 0.361 & 0.004 & 0.145 & 0.036 & 0.972 & 17\\
Run-Ran, 2010        &      1 & 0.361 & 0.004 & 0.147 & 0.031 & 0.965 & 19\\
Qiu, 2011            &      1 & 0.362 & 0.004 & 0.119 & 0.038 & 0.965 & 17\\
\rowcolor{lightsteelblue}
Zeng, 2013           &      1 & \tb{0.215} & 0.004 & 0.134 & \tb{0.023} & 0.708 & 40\\
\rowcolor{lightsteelblue}
Liu, 2012            &      2 & 0.361 & \tb{0.005} & \tb{0.170} & 0.030 & 0.963 & 19\\
L\"u, 2011*          &      2 & 0.361 & 0.005 & 0.148 & 0.031 & 0.965 & 19\\
Zeng, 2014           &      2 & 0.362 & 0.004 & 0.129 & 0.029 & 0.964 & 19\\
3hybrid1             &      3 & 0.249 & 0.004 & 0.152 & 0.025 & 0.904 & 30\\
3hybrid2             &      3 & 0.249 & 0.004 & 0.151 & 0.026 & 0.912 & 29\\
\end{tabular}
\end{ruledtabular}
\caption{Evaluation of the recommendation methods on the Amazon dataset ($L=50$).}
\label{tab:am}
\end{table*}

\section{Results}
\label{sec:results}
Results of the reviewed methods under triple data division are shown in Tables \ref{tab:ml}, \ref{tab:nf}, and \ref{tab:am}, respectively. For each method, we report the number of free parameters (recall that we choose the minimal ranking score as the criterion for parameter choice), ranking score $r$, precision $P(L)$, recall $R(L)$, list intra-similarity $I(L)$, list Hamming distance $H(L)$, and item novelty $N(L)$. The first thing to note is that related metrics generally give consistent results: a recommendation method with good ranking score has high precision and recall and a recommendation method with low novelty score has low intra-similarity and high Hamming distance. One can note that the most-popularity favoring methods do not achieve particularly high accuracy. The relation is even stronger for the most diversity-favoring methods (HeatS and Guo, 2012) that fail in recommendation accuracy.

A comparison of these results with those obtained using double data division (not shown here) is instructive in explaining the difference between these two approaches. The difference between the ranking score obtained in double and triple division is largest for the sparse Amazon data where it is around $0.02$ for all evaluated methods. This is because the risk of over-fitting and high ``apparent'' performance is particularly relevant for data that are sparse and/or limited in size. By contrast, results obtained with double and triple division differ by less than $0.002$ for almost all methods when evaluated on the Netflix data. Secondly, some methods that are claimed to be able to solve the cold-start problem seem not to work well with the triple division. For instance, ``Qiu, 2013'' improves the recommendation accuracy for small-degree items with double data division but once the triple division is employed, item novelty does not reach particularly low values. $N(L)$ shows that ``Qiu, 2011'', which is also said to mitigate the cold start problem, does not tend to recommend particularly less popular items than other methods. Finally, many methods that are claimed to outperform the classical ProbS-HeatS hybrid are found to have lower accuracy than the hybrid method, such as ``Guo, 2012'', ``Qiu, 2013'', ``Liu, 2011*'', and ``Zeng, 2013''. This can be due to a strong dependence of the optimal algorithm parameters on the data and simultaneously also a strong dependence of the method performance on the parameters' values. This is an important finding which demonstrates the importance of triple data division and the need to use it in any future research of network-based recommendation.

Among the parameter-free methods, ``Blattner, 2010'' outperforms ProbS for both the Movielens and Netflix data by achieving higher precision and recall and simultaneously lower intra-list similarity and novelty score and higher Hamming distance. The enhanced diversity of the method's recommendations is due to the fact that along with the probabilistic spreading it also employs the heat spreading. For one-parameter methods, ``Zhou, 2009'' seems to be the most accurate (though its diversity is not very good). Only slightly behind in accuracy, yet better in diversity is ``L\"u, 2011''. For two-parameter methods, ``Zeng, 2014'' seems to be the best in both accuracy and diversity. Comparing methods with different number of parameters, ``Zeng, 2014'' outperforms both ``Blattner, 2010'' and ``Zhou, 2009''. At least some methods with more parameters do not suffer from over-fitting and can produce better results than methods with fewer parameters. However, computational complexity of the parameter learning process grows in general quickly with the number of parameters (the simplest multidimensional search where the search range for each of $n$ parameters is divided in $l$ points requires to compute the recommendations and evaluate their performance for $l^n$ different parameter settings). If two parameters are of similar recommendation performance, the one with fewer parameters is therefore to be strongly favored.

Note that while both the Movielens and Netflix dataset are rather dense and consequently most of the methods perform similarly reasonably well on them, the Amazon dataset is very sparse (its density is two orders of magnitude lower) which manifests itself in low recommendation accuracy of all methods. The low precision values found for the Amazon data, for example, can be put into perspective by comparing with the precision of random recommendations as in~\cite{zhou2010solving}. The best-performing method ``Liu, 2012'', for example, is 170-times more precise than random recommendations. The ranking score achieved by the methods on the Amazon data is bad mainly due to the fact that only a limited number of items can be reached in two diffusion steps; the unreachable items obtain identical zero recommendation score and the probe items among them then contribute to a high ranking score. Methods with more than two diffusion steps consequently achieve much better ranking score which, however, does not automatically translate into high precision and recall (see ``Qiu, 2011'' which on the Amazon data has the second-best ranking score and at the same time the worst precision).

Based on the result tables, one can conclude that the following features generally positively contribute to the performance of a network-based method: i) Modifying the initial configuration based on the degree of selected items, ii) Amplifying the resource on users with higher similarity to the target user, iii) Considering multiple-step diffusion. We have thus attempted to design an effective recommendation algorithm by combining these three ingredients in a three-parameter recommendation method (see the description of 3hybrid1 and 3hybrid2 in Section~\ref{sec:methods}). The results show that the new methods can indeed outperform most of the other methods included in this review (for the Movielens data, 3hybrid1 is the best of all methods in all three accuracy-focused metrics). However, the improvement gain seems rather small which suggests that the potential for improvements through combination of various methods is limited when three- and more-parameter combinations are considered. The increased computational complexity of the many parameter methods (see above) is an additional argument for limiting ourselves to methods with two parameters or less.

Compared with traditional recommendation algorithms such as collaborative filtering, an important advantage of network-based recommendation algorithms is in their formulation using a propagation process on a network. In this way, one can avoid constructing the user or item similarity matrix which is usually very big because the number of users and items in real systems is high. Due to the low density of these real networks, the propagation description of network-based methods requires us to only store the score of the items that are reachable in a few random walk steps on the network, which reduces the computational complexity dramatically. On the other hand, there are also network-based recommendation algorithms such as ``Blattner, 2010'', ``Qiu, 2013'' and ``Liu, 2011*'' that cannot be effectively described by a propagation process and the described computational advantage thus does not apply to them.

\section{Beyond algorithms}
\label{sec:beyond}

\subsection{Mutual feedback between recommender systems and real system evolution}
A recommender system is an effective tool to find the most relevant information for online users. At the same time, it may significantly influence the distribution of item popularity. This is due to the fact that recommendation guides people's choice, which further influences subsequent recommendations and hence the choice of others. The influence of this mutual feedback between users and recommender systems is hence amplified with successive recommendations. Recommendation algorithms may have very different long-term influence on the user-item network evolution. For example, if a recommendation algorithm always recommends popular items, gradually only the most popular items survive, causing the market to be dominated by these items. On the other hand, if a recommendation algorithm tends to recommend less popular items, item popularity may with time become more homogeneous.

\begin{figure*}
\centering
\includegraphics[scale=0.5]{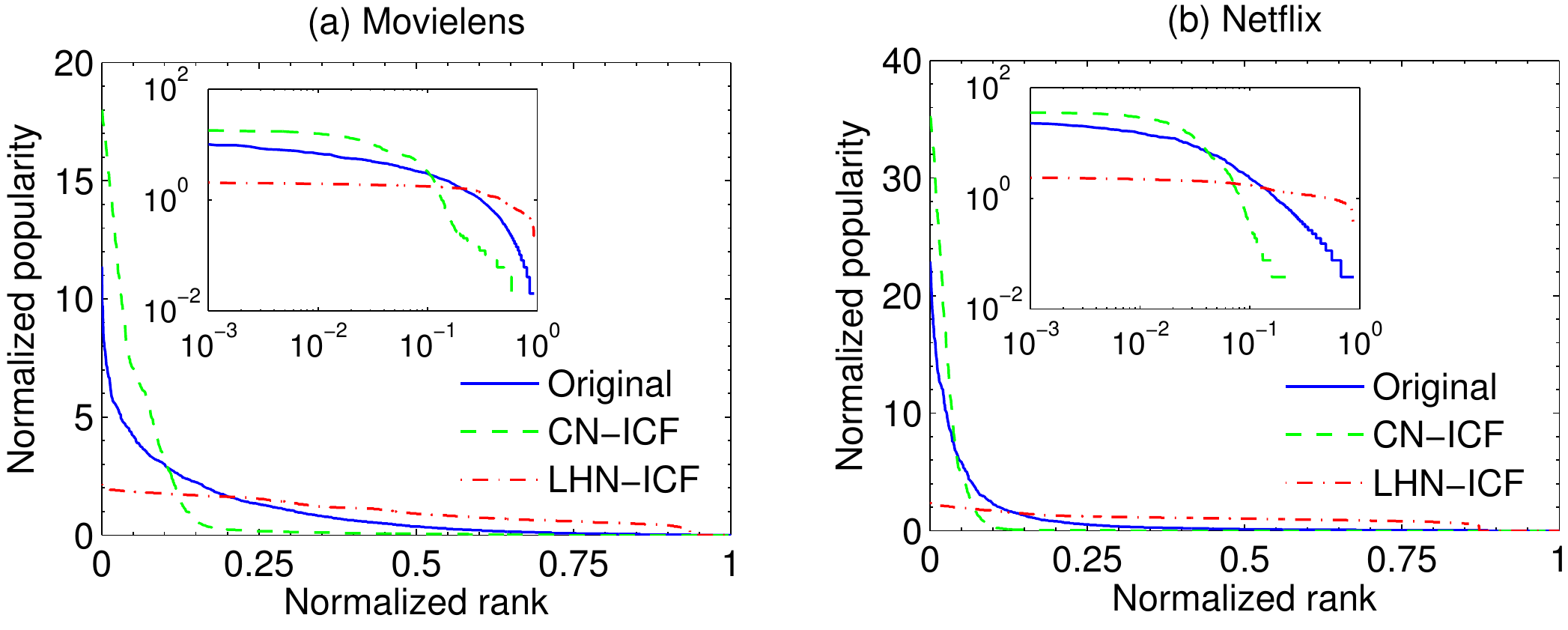}
\caption{The normalized item popularity versus the normalized item rank in (a) Movielens and (b) Netflix. The same plots in log-log scale are shown in the respective insets. We compare here the original input data with the data repeatedly influenced by the use of item-based collaborative filtering (ICF) with two different similarity metrics: common neighbors (CN) and Leicht-Holme-Newman (LHN) similarity. (Adapted from~\cite{zeng15mutual}.)}
\label{fig:feedback}
\end{figure*}

To tackle this problem, \cite{zeng15mutual} develops a model to study the long-term co-evolution of recommendations generated by a recommender system and the user-item network representing the choice of users. Starting from the user-item bipartite networks constructed from real observed data, the model rewires links by assuming that users seek for new content and in doing so, they with some probability accept recommendations generated by a collaborative filtering algorithm or otherwise choose an item at random (or, alternatively, proportionally to item popularity). The network evolves to a quasi-stationary state in which the item degree distribution does not further change its overall shape. The authors stress that we need to distinct between two different aspects of item diversity: short-term diversity which relates to items recommended to users on the basis of given data and long-term diversity which relates to the distribution of item popularity upon a long-term use of a given recommendation method. Contrary to the common belief that recommendation widens our information horizons, this model shows that the whole information ecosystem results in a situation where a few items enjoy extraordinarily high levels of user attention (see Fig.~\ref{fig:feedback} where upon the use of a popularity-favoring common neighbors similarity metric in recommendations, the distribution of item popularity becomes more uneven). In other words, long-term diversity is very low. Moreover, a strong hysteresis effect is present which implies that once formed, the state of uneven item popularity is difficult to be reverted, even when a diversity-supporting recommendation method is used. While enhancing the diversity of recommendations typically results in lowering their accuracy, \cite{zeng15mutual} suggests a solution to this challenge by demonstrating that there is a possible favorable trade-off between recommendation accuracy and long-term diversity.

\subsection{Crowd-avoidance}
Previously described recommendation methods attempt to improve the recommendation performance (roughly speaking, accuracy and diversity) by modifying the spreading process that is used to compute the resulting recommendation scores. It is also possible to take recommendation scores produced by a given method and act on them in such a way that the diversity of resulting recommendations increases. This has been studied in~\cite{ziegler2005improving} where the authors aim to increase the diversity of items in each user's recommendation list and in~\cite{zhang2008avoiding} where a trade-off between optimizing the user-item matching and maintaining high average item dissimilarity is studied. A different class of approaches is based on combining the usual ranking of items by the predicted score with other alternative rankings such as the rating by overall item popularity, average item rating, or rating variance. As shown in~\cite{adomavicius2012improving}, thus-achieved diversity gain is often big in comparison with the corresponding precision loss. A very different approach is based on limiting the number of users to whom any individual item can be recommended (crowd avoiding)~\cite{gualdi2013crowd}. While the introduction of an ``occupation constraint'' on items obviously has the potential to improve recommendation diversity, recommendation accuracy can be improved in this way too. The authors hypothesize that the improvement of recommendation accuracy is due to the constraint effectively removing the popularity bias that is innate to many recommendation methods. To support their claims, they propose a simple artificial model where users' recommendation lists share some bias and show that applying an occupation constraint improves recommendation accuracy in the same way as in the real-world datasets. In this section, we describe the crowd avoidance approach in detail and significantly extend the results presented in~\cite{gualdi2013crowd}.

We assume that a recommendation method (any method, not necessarily one of the network-based methods reviewed here) produces a ranked list of items for each user. The rank of item $\alpha$ in the recommendation list of user $i$ is denoted as $r_{i\alpha}$ (the smaller, the better). The simplest way to apply the crowd-avoidance principle is to constrain the number of users to whom each individual item can be recommended to $m$ (so-called maximal occupancy). When $m$ is small, selecting the top-ranked item for each user is likely to result in some items being assigned to more than $m$ users. Since this is not allowed, we choose a reverse approach: to assign users with items that are as good ranked as possible whilst obeying the occupancy constraint. Denoting the item assigned to user $i$ as $\beta(i)$, the number of users whom item $\alpha$ has been assigned is $n_{\alpha} := \lvert \{i: \beta(i)=\alpha\}\rvert$. Our task is thus to minimize
\begin{equation}
\label{objective}
\sum_{i=1}^U r_{i\beta(i)}
\end{equation}
which is our objective function with the set of constraints
\begin{equation}
\forall\alpha:\quad n_{\alpha}\leq m
\end{equation}
that apply to each item individually. This optimization problem can be solved locally by going over all users in a random order and assigning each user with the Most-Preferred Object (MPO) that is currently assigned to less than $m$ users. If several items are to be assigned to each user, one repeats the cycle over users several items until the required number of items per user is reached. This approach has the advantage of being computationally efficient and easy to implement but it obviously yields sub-optimal solution, especially when the occupancy constraint $m$ is low and one often has to assign low-ranked items to users.

The alternative is to search for the globally optimal assignment that minimizes Eq.~(\ref{objective}) and respects the constraints. As already reported in~\cite{gualdi2013crowd}, the exact solution is provided by the Hungarian Algorithm (HA) which solves this task in polynomial time~\cite{kuhn1955hungarian}. The disadvantage of the exact approach is that it originally applies to the case $m=1$ (each item is assigned to at most one user). While it is conceptually easy to modify the algorithm to $m>1$ by creating $m$ artificial copies of each item, this results in increasing the time and memory complexity of the algorithm. As a result, the original crowd-avoidance paper only studies global optimization for $m\leq12$. To overcome this problem, we present here two approximate approaches to global optimization. Simulated Annealing (SA) is a well-known optimization algorithm for high-dimensional optimization problems~\cite{van1987simulated} which is eminent for its ability to escape from local minima by sometimes accepting changes that worsen the solution (see below for implementation details). For comparison, we also apply a strictly greedy local optimization technique where only favorable changes of the solution are accepted.

\begin{figure}
\centering
\vspace*{6pt}
\includegraphics[scale=0.35]{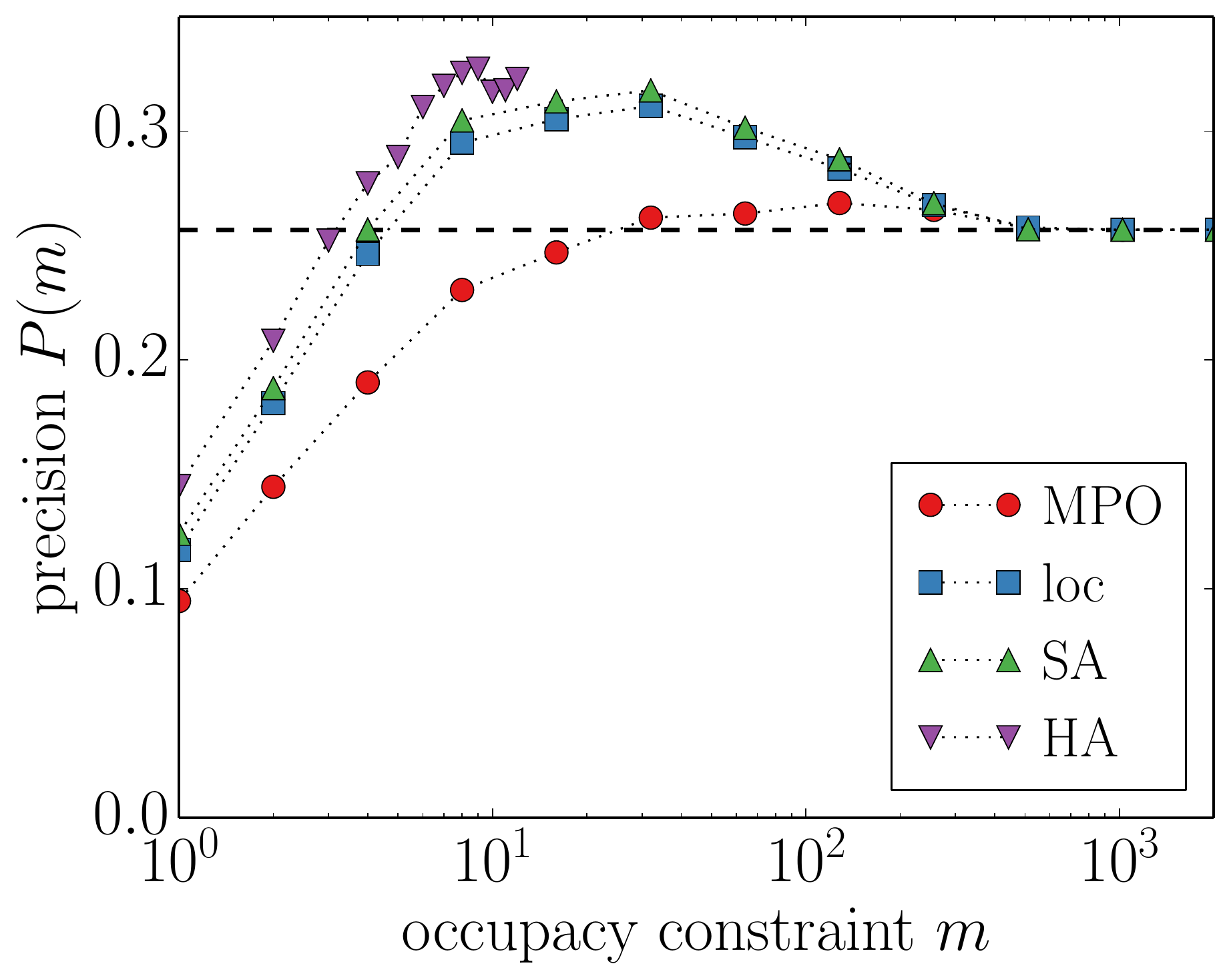}
\caption{A comparison of the four assignment methods (Most-Preferred Object, local optimization, Simulated Annealing, and Hungarian Algorithm) that respect the general occupancy constraint when $L=1$ (\ie, one item is assigned to each user).}
\label{fig:crowd}
\end{figure}

Results in this section are based on a small subset of the Netflix data with 2,000 users and 2,000 items that have been chosen at random. After the usual thresholding procedure, the dataset contains 592,995 user-item links. As in~\cite{gualdi2013crowd}, we assume here that only one item is to be recommended to each user. In addition to precision, we use the effective number of recommended items
\begin{equation}
\neff = \frac{(LU)^2}{\sum_{\alpha=1}^I n_{\alpha}^2}
\end{equation}
to demonstrate the effect of crowd-avoidance on recommendation diversity. Here $L$ is the recommendation list length (the number of items recommended to every user). This metric's extreme values are straightforward: (1) When all items are recommended equally often, $n_{\alpha} = UL / I$ and $\neff = I$ (the highest possible value), (2) When the same $L$ items are recommended to all users, $\neff=L$ (the smallest possible value). Figure~\ref{fig:crowd} shows the dependence of recommendation precision on the occupancy constraint $m$ for the four described implementations of crowd-avoidance. While the computationally most expensive SA approach is able to improve recommendation precision much more than the computationally cheapest MPO approach, loc and SA also offer significant improvements with less computation than SA. The methods' relative improvements with respect to the benchmark assignment without crowd-avoidance ($m\to\infty$) are 5\%, 21\%, 24\%, and 27\% for MPO, loc, SA, and HA, respectively. The values of $\neff$, while varying strongly with $m$, differ little among the methods. There is an important difference, though, that recommendation precision achieved with well-performing methods peaks at lower values of $m$ (\ie, stronger crowd-avoidance) than it does for MPO. As a result, $\neff$ achieved at the peak of precision with MPO is 3.9-times higher than in the unconstrained case (when $\neff=4.7$), while the ratio is 14 for loc and SA and it exceeds 50 for HA. We can conclude that crowd-avoidance indeed helps to increase recommendation precision and, even more, recommendation diversity. The two new assignment methods, local optimization (loc) and simulated annealing (SA), are computationally more effective in achieving this goal than the previously suggested exact optimization with the Hungarian algorithm (HA).

We finally briefly describe the implementation details of the simulated annealing algorithm used to obtain the aforementioned results. Items are initially assigned to users on the basis of the MPO assignment. This configuration then evolves in so-called macro steps that consist of $100U$ micro steps each (a smaller multiple of $U$ can be used to reduce the simulation time at the expense of some accuracy loss). In every micro step, we choose two users $i$ and $j$ at random and compute the rank change
\begin{equation}
\Delta R:=r_{i\beta(j)} + r_{j\beta(i)} - (r_{i\beta(i)} + r_{j\beta(j)})
\end{equation}
that would follow from exchanging the objects between these two users. In local optimization, steps with $\Delta R < 0$ are accepted and those with $\Delta R > 0$ are rejected. In simulated annealing treats the steps with $\Delta R < 0$ the same and steps with $\Delta R > 0$ are accepted with the probability $\exp(-\Delta R / T)$ where $T$ is the temperature that is initially set high (we use $10^4$ in our simulations which is more than any possible rank change in a system with 2000 items) and then slowly decreases (we multiply $T$ with $0.95$ after each macro step). Simulation ends after there is no change encountered in a whole macro step; since SA can take a long time to meet this criterion, we also stop after 1000 macro steps have been realized. In addition to micro-steps based on two-user changes, one could also employ one-user changes where we attempt to assign a randomly chosen with a randomly chosen item. Including one-user changes in our simulations does not result in improving the results significantly.

\section{Discussion}
\label{sec:discussion}
In this review, we compare for the first time a comprehensive set of network-based recommendation methods. We believe that such a comparison is greatly needed as it allows one to assess the performance of various methods and, in particular, computational complexity induced by the number of parameters that enter in the performance-optimization process. To make a fair comparison of methods possible, we thoroughly use the triple division of the input data where the first 80\% of the data is used to determine the optimal parameter values by comparing the obtained recommendations with hidden 10\% of the data. The resulting parameter values are then used to compute recommendations that are finally compared with the last remaining 10\% of the data. As explained in Section~\ref{sec:triple}, this approach avoids the potential problem of method over-fitting and the resulting overestimation of the method's performance. As we discuss in Section~\ref{sec:results}, these dangers are not only illusory---they have probably contributed to reported superior performance of methods that in our comparison actually do not perform better than the others.

In conclusion, we would like to discuss two research topics in this field that are of particular importance in our view. First, all recommendation methods presented here have been evaluated (both in original papers as well as here) with random division of the input data where a randomly chosen part of the data is moved to the so-called probe. However, the probe consisting of the most recent links would reflect the use of recommendation in reality much better than a random probe. After all, the goal of recommendation is to help choose the potential \emph{future} likes and interests for individual users. Our preliminary work shows that upon the time-based probe, recommendation performance of all methods importantly decreases. The performance decreases remains even when the new items that first appear in the probe set are excluded from performance evaluation. This is understandable because while the methods discussed in this review examine intensively the personalized patterns of interest of individual users, none of them examines the temporal patterns of content popularity~\cite{szabo2010predicting,medo2011temporal,zeng2013trend}. This phenomenon calls for recommendation methods that take time into account and thus better reflect the real temporal evolution of systems.

Second, the question of mutual influence between information filtering algorithms and the systems in which they act, although considered by some empirical works~\cite{basuroy2003critical,salganik2006experimental}, is still largely unexplored. For instance, the model introduced in~\cite{zeng15mutual} assumes that all users accept the recommendation with the same probability. One should also consider a scenario where experienced users search for items on their own and thus depend less on recommendation. The input provided by the experienced users can prove instrumental for the evolution of the system. It would be interesting to monitor the recommendation accuracy during the system's evolution and couple it with the rate at which users accept recommendations.

\begin{acknowledgments}
This work was supported by the Swiss National Science Foundation Grant No. 200020-143272 and by the EU FET-Open Grant No. 611272 (project Growthcom). A.Z. is supported by the Young Scholar Program of Beijing Normal University (grant No. 2014NT38). We thank Yi-Cheng Zhang for stimulating discussions on recommendation and its economic and societal impacts.
\end{acknowledgments}

\bibliographystyle{elsarticle-num}
\bibliography{rec_rev_ref}

\end{document}